\documentclass[10pt,onecolumn,english] {IEEEtran}
\ifCLASSINFOpdf
\usepackage[pdftex]{graphicx}
\usepackage{multirow}
\usepackage{textpos}
\usepackage{amsmath}
\usepackage{amsfonts}
\usepackage{amssymb}
\usepackage{array}
\usepackage{color}
\usepackage{colortbl}
\usepackage{amssymb}
\usepackage{setspace}
\usepackage{multicol,lipsum}
\usepackage{cuted}
\usepackage{amsthm}

\newtheorem*{remark}{Remark}
\newtheorem*{discussion}{Discussion}

\else
\fi

\begin{document}
%
%\begin{frontmatter}
%
\title{On The Use of Conjugate Teager-Kaiser Energy Operators with Matched Filters}
\author{Jean-Philippe~Montillet, \IEEEmembership{Senior Member,~IEEE} 
\thanks{J.P.~Montillet is with the School of Engineering, Electronics and Signal Processing (ESPLAB), Ecole Polytechnique F\'ed\'erale de Lausanne (e-mail: jean-philippe.montillet@epfl.ch)}}
%
%\author{Dr. Jean-Philippe~Montillet, SMIEEE ,  email:jean-philippe.montillet@epfl.ch}
%> for one column

%{Akaike criterion based on a bivariate normal distribution for frequency estimates of two signals closely spaced in Frequency}
%\title{The Generalization of the Decomposition of Functions by Energy Operators}
% and the Application to the Linear Partial Differential Equations}
%

\maketitle

\markboth{Working Paper July 2016}
%{IEEE Trans. on Information Theory ,~Vol.~x, No.~x, xxx~ xxxx}%
{Montillet \MakeLowercase{}: On The Use of Energy Operators with Matched Filters}
%{Shell \MakeLowercase{\textit{et al.}}: Bare Demo of IEEEtran.cls for Journals}
\doublespacing
%
%\maketitle
\begin{abstract}
\boldmath{
This work presents the proof of concept of using energy operator theory based on the conjugate Teager-Kaiser energy operators in matched filters for signal detection in multipath fading channels. To do so, we consider  signals in the space $\mathcal{S}(\mathbb{R})$ a subspace of the Schwartz space $\mathbf{S}^-(\mathbb{R})$ in order to approximate the received signal. These functions obey to some specific properties as shown in this work.  It allows to decompose the received signals into subchannels with their own Signal-to-Noise-Ratio.}
\end{abstract}

%The presented work is not based on the current main stream research work surrounding this operator. 
%
\section{Introduction}\label{Introduction}
In wireless channels, a signal may go through different fading paths due to reflections from obstacles. A radio channel can thus consist of many copies of originally transmitted signals having different amplitudes, phases, and delays.
Since each path has a different length and the signals travel at the same speed, the signal arrival times for the paths differ. If the signal components arrive more than the duration of one chip apart from each other, a rake receiver can be used to resolve and combine them \cite{Prasad}. %The basic functions of rake receiver are channel delay estimation for multipath components, rake receiver finger allocation, descrambling and despreading operations, adaptive channel estimation, and Maximal-Ratio Combining \cite{Prasad}.
% The rake receiver uses a multipath diversity principle. It is widely used in wireless  telecommunications \cite{Stuber} and in particular in mobile communications in  Wide band Code Division Multiple Access (WCDMA) and in particular Direct Sequence CDMA due to the wider bandwidth and hence large resolution in order to separate the different paths received at the antenna \cite{Prasad} \cite{Miller}. 
We assume a signal transmitted through an  impulse response channel defined \cite{Proakis}:
\begin{equation}\label{ImpulseChanneldef}
r(t) = \sum_{l \in \mathbb{Z}^+} a_l(t)  e^{j \phi_l (t)} f(t-\tau_l) + \eta(t)
\end{equation}
where $\phi_l(t)$ is the phase, $a_l(t) $ the amplitude, $f(t)$ the transmitted signal and $\eta(t)$ a white noise. The stochastic property of the noise follows a Gaussian distribution.% $\mathcal{N}(0,\sigma^2_0)$.
\\Several decades ago, the energy operator was introduced for the first time as the Teager-Kaiser energy operator \cite{Kaiser90} in signal processing specifically for the detection of transient signals \cite{Dunn} and also filtering modulated signals \cite{Bovik93}. A few years later the family of Teager-Kaiser energy operators was defined by \cite{Maragos1995}. Since, many applications have been found such as speech analysis\cite{Dunn}, transient signal detection\cite{Kandia2006}, image processing \cite{Cexus},  optic \cite{Salzenstein}, and localization \cite{Schasse}. A few years ago,  several works (e.g, \cite{JPMontillet2010}, \cite{JPMontillet2013} and \cite{JPMontillet2014}) redefined the initial Teager-Kaiser energy operator such as $\Psi_2^-$ in order to introduce the conjugate operator $\Psi_2^+$ and rewriting the wave equation with these two operators. Subsequently, \cite{JPMontillet2013} defines the families of energy operator $( \Psi_k^-)_{k\in\mathbb{Z}}$ and $( \Psi_k^+)_{k\in\mathbb{Z}}$ in order to decompose the successive derivatives of a finite energy function $f^n$ ($n$ in $\mathbb{Z}^+ -\{0,1\}$) in the Schwartz space $\mathbf{S}^-(\mathbb{R})$. 
%
%\
%
Because of the formulation of Lemma $1$   together with its proof in  \cite{JPMontillet2013}, one can write  $\partial_t^k f^n$ ($k$ in $\mathbb{Z}^+$, $k>0$, $n$ in  $\mathbb{Z}^+$, $n>1$) with a sum of the energy operators $( \Psi_k^+)_{k\in\mathbb{Z}}$. 
\\ The focus of this work is the proof of concept on how the energy operators $( \Psi_k^+)_{k\in\mathbb{Z}}$ can be used to detect and decorrelate some of the reflected paths in a multipath fading channel based on the redefinition of the Signal-to-Noise-Ratio (SNR) with those operators. In theory, the energy operators could then be used to define multipath fingers of a rake receiver  for specific signals in a subspace of  $\mathbf{S}^-(\mathbb{R})$. This subspace is defined as  $\mathcal{S}(\mathbb{R})$ in the following sections.
\\ In the next section, we recall some of the properties for functions belonging to the Schwartz space $\mathbf{S}^-(\mathbb{R})$. In Section \ref{section3}, the subspace $\mathcal{S}(\mathbb{R})$ is defined in order to justify the introduction of the special impulse response channel for the signals belonging to this space. Finally, we show  how the multipath rays can be predicted when using the energy operators and their derivatives based on the closed-form formula of the special impulse response channel. The last section is the definition of a matched filter with the help of energy operators.
%   
%
%\section{Some Definitions on the Schwartz Space $\mathcal{S}^-{\mathbb{R}\} $}\label{section2}
%
\section{Preliminaries on the Schwartz Space $\mathbf{S}^{-}(\mathbb{R})$}\label{preliminariesSection}
Let us define $f^n$ for any $n$ in $\mathbb{Z}^+-\{0\}$. $f^n$ is  supposed to be a smooth real-valued and finite energy function, and in the Schwartz space $\mathbf{S}^{-}(\mathbb{R})$ following the definition \cite{Reed}:
\begin{eqnarray}\label{SRRRRR}
\mathbf{S}^{-}(\mathbb{R}) &=&\{f \in \mathbf{C}^{\infty}(\mathbb{R}), \hspace{0.5em} {sup}_{t<0} |t^k||\partial_t^j f(t)|<\infty, \nonumber \\
& & \hspace{0.5em}\forall k \in \mathbb{Z}^+, \hspace{0.5em} \forall j \in \mathbb{Z}^+ \}
\end{eqnarray}
with $\mathbf{C}^{\infty}(\mathbb{R})$ the class of infinitely differentiable functions with value in $\mathbb{R}$. Sometime $f^n$ can also be analytic if its development in Taylor-Series is relevant to this work. The choice of $f^n$ (for any $n$ in $\mathbb{Z}^+-\{0\}$) in the Schwartz space $\mathbf{S}^{-}(\mathbb{R})$ is based on the work developed in \cite{JPMontillet2013}. Here, we are dealing with multiple integrations or derivatives of $f^n$ when applying the energy operators $({\Psi}_{k}^{-})_{k\in\mathbb{Z}^+}$. $({\Psi}_{k}^{+})_{k\in\mathbb{Z}^+}$ is defined \cite{JPMontillet2013}:
\begin{equation}\label{DefinitionOperator}
{\Psi}_{k}^{+/-}(.) = \partial_t . \times \partial_t^{k-1} . \pm . \times \partial_t^k .
\end{equation}
Note that the Teager-Kaiser energy operator was initially defined in \cite{Kaiser90} in thespecific case of $k$ equal $2$ with the definition ${\Psi}_{k}^{+}(.)$ in Equation \eqref{DefinitionOperator}. In the following, let us call the set $\mathcal{F}(\mathbf{S}^{-}(\mathbb{R}),\mathbf{S}^{-}(\mathbb{R}))$ all functions/operators defined such as $\gamma:$ $\mathbf{S}^{-}(\mathbb{R})$ $\rightarrow$ $\mathbf{S}^{-}(\mathbb{R})$. Let us recall some definitions and important results given in \cite{JPMontillet2013}.
%
%\vspace{1.0em}
%
From \cite{JPMontillet2013} and \cite{JPMontillet2014}, the properties of the energy operators can be summarized as:
\begin{itemize}
\item Quadratic  form in the set of functions $\mathcal{F}(\mathbf{S}^-(\mathbb{R}),\mathbf{S}^-(\mathbb{R}))$
\item Bilinearity
\item Derivative chain rule property: $\partial_t {\Psi}_{k}^{-/+}(f) = {\Psi}_{k+1}^{-/+}(f) +{\Psi}_{k-1}^{-/+}(\partial_t f) $, for $k$ in $\mathbb{Z}^+$
\end{itemize}
All those properties are shown in \cite{JPMontillet2013} (e.g, Section $2$ and $3$). Let us recall some definitions and previous results.
\vspace{1.0em}
\\$\bold{Definition}$ $1$ \cite{JPMontillet2013}: \emph{for all $f$ in $\mathbf{S}^{-}(\mathbb{R})$, for all $v\in\mathbb{Z}^+-\{0\}$, for all  $n\in\mathbb{Z}^+$ and $n>1$, the family of operators $(\Psi_k)_{k \in \mathbb{Z}}$ (with $(\Psi_k)_{k \in \mathbb{Z}}$ $\subseteq$ $\mathcal{F}(\mathbf{S}^{-}(\mathbb{R}),\mathbf{S}^{-}(\mathbb{R}))$) decomposes $\partial_t^v$$f^n$ in $\mathbb{R}$, if it exists $(N_j)_{j\in \mathbb{Z}^+ \cup \{0\}}$ $\subseteq$ $\mathbb{Z^+}$,  $(C_i)_{i=-N_j}^{N_j}$ $\subseteq$ $\mathbb{R}$, and it exists $(\alpha_j)$ and $l$ in $\mathbb{Z^+}\cup\{0\}$ (with $l<v$) such as  $\partial_t^v$$f^n = \sum_{j=0}^{v-1} \big(_{j}^{v-1} \big) \partial_t^{v-1-j} f^{n-l} \sum_{k=-N_j}^{N_j} C_k \Psi_k(\partial_t^{\alpha_k}f)$.}
%
%\vspace{1.0em}
%
%%
%In addition, one has to define $\mathbf{s}^{-}(\mathbb{R})$ as:
%%
\vspace{0.5em}
\\ An important result shown in \cite{JPMontillet2013} is: 
\vspace{0.5em}
\\$\bold{Lemma}$ $0$ \cite{JPMontillet2013}: \emph{for $f$ in $\mathbf{S}^{-}(\mathbb{R})$, the family of energy operators ${\Psi}_{k}^{+}$ ($k=\{0,\pm 1,\pm 2,...\}$) decomposes the successive derivatives of the $n$-th power of $f$ for $n\in\mathbb{Z}^+$ and $n>1$.} 
\begin{proof}
The full proof is given in \cite{JPMontillet2013}.
\end{proof}
%\%vspace{1.0em}
%
%\\$\bold{Theorem}$ $0$ \cite{JPMontillet2013}: for $f$ in $\mathbf{s}^{-}(\mathbb{R})$, the families of DEO ${\Psi}_{k}^{+}$ and ${\Psi}_{k}^{-}$ ($k=\{0,\pm 1,\pm 2,...\}$) decompose uniquely the successive derivatives of the $n$-th power of $f$ for $n\in\mathbb{Z}^+$ and $n>1$. 
%
\vspace{0.5em}
By definition if $f^n$ is analytic, there are ($p$,$q$) ($p >q$) in $\mathbb{R}^2$ such as $f^n$ can be developed in Taylor Series \cite{Kreizig2003}:
\begin{eqnarray}\label{f2eq}
f^n(p) &=& f^n(q) + \sum_{k=1}^\infty \partial_t^k f^n(q) \frac{(p-q)^k}{k!} \nonumber \\
\end{eqnarray}
Let us define for $n$ in $\mathbb{Z}^+-\{0\}$, for $f^n$ in $\mathbf{S}^{-}(\mathbb{R})$ and finite energy, $\mathcal{E}(f^n)$  the energy function defined for ($\tau$,$q$) ($q< \tau$) in $\mathbb{R}^2$ such as:
\begin{equation}\label{EnergyfunDefine02}
\mathcal{E}(f^n(\tau)) = \int_q^{\tau} |f^n(t)|^2dt < \infty
\end{equation}
\vspace{0.5em}
\\$\bold{Proposition}$ $1$ \cite{JPMontillet2013} \emph{If for any $n$ in $\mathbb{Z}^+$, $f^n$ in $S^−(\mathbb{R})$ is analytic and finite energy; for any $(p,q)$ in $\mathbb{R}^2$ (with $p > q$) and $\mathcal{E}(f^n)$ in $S^−(\mathbb{R})$ is analytic, then $\mathcal{E}(f^n(p))$ is a convergent series.}
\begin{proof}
The full proof is given in \cite{JPMontillet2013}.
\end{proof}
%
%\newline $\bold{Proposition}$ $1$ \cite{JPMontillet2014}: If for any $n$ in $\mathbb{Z}^+$, $f^n$ in $\mathbf{S}^{-}(\mathbb{R})$ is analytic and finite energy; for any ($p$,$q$) in $\mathbb{R}^2$ (with $p>q$) and $\mathcal{E}(f^n)$ in $\mathbf{S}^{-}(\mathbb{R})$ is analytic, then $\mathcal{E}(f^n(p))$ is a convergent series.
%
\vspace{0.5em}
In the following sections, we are interested in the specific application of $f(t)$, a signal transmitted through a multipath channel with some specific properties. 
\section{Channel Impulse Response Approximation for signals in $\mathcal{S}(\mathbb{R})$}\label{section3}
This section defines the approximation of classical type of multipath fading channel, precisely a modified impulse response channel based on Equation \eqref{ImpulseChanneldef}, together with the type of signals that can be used in this approximation.
%Note that we consider the phase slowly varying ($\bar{\phi_l(t)}| \simeq 0$).
%
\subsection{ Channel Impulse Response Approximation}
\vspace{0.5em}
$\mathbf{Definition}$ $2$ \emph{(approximation of multipath fading channel): if $g$  is a function of time $t$ from $\mathbb{R} \rightarrow \mathbb{R}$ transmitted through the multipath cahnnel $h(t) = \sum_{l \in \mathbb{Z}^+} \rho_l \partial_t^l \delta(t-\tau_l) $ (for all $t$ $\in$ $\mathbb{Z}^+$), $\rho_l$ in $\mathbb{R}$, $\delta$ the Dirac function \cite{Kreizig2003}. Then the received signal $r(t)$ is approximated by the successive derivatives of $f^n(t)$, with $f^n$ is in $S(\mathbb{R})$ and the associated signal $f^n(t)$ in $\mathbb{R}$ for all $t$ $\in$ $\mathbb{Z}^+$, $n \in \mathbb{Z}^+$, $n>0$. The channel impulse approximation is then defined as:}% received signal $r(t)$ in additive white Gaussian noise (AWGN) and multipath fading channel can be written such as:
\begin{eqnarray}\label{receivedsignaldef}
%r(t) = \sum_{k \in \mathbb{Z}^+} \sum_{L\in \mathbb{Z}^+-\{0\}}\rho_k \partial_t^L f(t-\tau_k) + \eta(t)
r(t) &=& h(t) \otimes g(t) + \eta(t) \nonumber \\
r(t) &=& \sum_{l \in \mathbb{Z}^+} a_l(t)  e^{j \phi_l (t)} g(t-\tau_l) + \eta(t) \nonumber \\
r(t) &\simeq & \sum_{l \in \mathbb{Z}^+} \sum_{k \in \mathbb{Z}^+} \sum_{n \in \mathbb{Z}^+, n>1}\beta_k(l) \partial_t^k f^n(t-\tau_l) \nonumber \\
\end{eqnarray}
\vspace{0.5em}
\emph{$\eta$ is the AWGN, $\tau_l$ the delay for each ray received by the antenna and $\beta_k$  the amplitude in $\mathbb{R}$}. $\otimes$ is the convolution operator. %The special multipath channel $h(t)$ is equal to $h(t) = \sum_{l \in \mathbb{Z}^+} \rho_l \partial_t^l \delta(t-\tau_l) $. $\delta$ is the Dirac function \cite{Kreizig2003}.
\vspace{0.5em}
\\ In other words, we define the projection of the received signal $r(t)$ onto the manifold defined by the basis $[\partial_t^k f^n(t)]_{k \in \mathbb{Z}^+,n \in \mathbb{Z}^+, n>1}$. 
%emphasizes that the derivatives of the  signal $f^n$ in the subset $\mathcal{S(}\mathbb{R})$, define the various paths (multipath) of reflected signals received by the antenna.
However, there are some limitations on the function $f^n$ describes with the following properties:
\vspace{0.5em}
\\$\mathbf{Property}$ $1$: \emph{ $f^n$ in $\mathcal{S}(\mathbb{R})$, and $\mathcal{S}(\mathbb{R}) \subsetneq \mathbf{S}^-(\mathbb{R})$, $n \in \mathbb{Z}^+$, $n>1$, $f^n$ is finite energy. For all $l \in \mathbb{Z}^+$,} 
\begin{itemize}
\item[1.] \emph{$\exists$ $l_0\in \mathbb{Z}^+$ such as for all  $l\in \mathbb{Z}$, $l>l_0$, $\partial_t^l f^n \sim 0$. }
\item[2.] \emph{for all $t$ $\in$ $\mathbb{Z}^+$, $\sum_{l\in \mathbb{Z}^+} \partial_t ^l f^n(t) <\infty $.}
\item[3.] \emph{$f$ is analytic and its Taylor series development is convergent for all $(t_1, t_2) \in (\mathbb{Z}^+)^2$, $t_1 < t_2$: $f^n(t_2) = f^n(t_1) + \sum_{l=0}^{l_0} \partial_t^l f^n(t_1) \frac{(t_2-t_1)^l}{l!}$.}
\end{itemize}
\vspace{0.5em}
\begin{remark}[1]
 In Property $1$, $[1] \rightarrow [2]$ with  $\sum_{l\in \mathbb{Z}^+} \partial_t ^l f^n(t) = \sum_{l =0}^{l_0} \partial_t ^l f^n(t)  $, and $\sum_{l\in \mathbb{Z}^+} \partial_t ^l f^n(t) < l_0 \times max_{l \in [0, l_0-1]} \partial_t ^l f^n(t) $. In other words, $Property$ $1$ guarantees  that there is no infinite number of rays being received at a given delay $\tau_l$. This condition is also called the Finite Impulse Response (FIR) \cite{Proakis}.
\end{remark}
%
%Using $\mathbf{Definition}$ $1$, $L$ is now a function of $k$ in equation \eqref{receivedsignaldef} such as:
%%
%\begin{equation}\label{receivedsignaldef}
%r(t) = \sum_{k \in \mathbb{Z}^+} \sum_{L=0}^{L_0(k)}\rho_k \partial_t^L f(t-\tau_k) + \eta(t)
%\end{equation}
%%
\begin{remark}[2]We could restrict $\mathcal{S}(\mathbb{R}) \subsetneq \mathcal{C}^{\infty}(\mathbb{R})$, but using a subspace of the Schwartz space $\mathbf{S}(\mathbb{R}^-)$ allows further developments with the energy operators due to the multiple derivatives or integrations \cite{JPMontillet2013}.
\end{remark}
\begin{remark}[3]
With $f^n$ following $1$ in $Property$ $1$, it then exists  $l_0\in \mathbb{Z}^+$ such as for all  $l\in \mathbb{Z}^+$, $l>l_0$, $\partial_t^l f \sim 0$. $Property$ $1$  guarantees that the basis  $[\partial_t^k f^n(t)]_{k \in \mathbb{Z}^+}$ is not infinite. Furthermore, if $\mathbf{V}$ is the vector space defined by the basis $[\partial_t^k f^n(t)]_{k \in \mathbb{Z}^+}$, $\mathbf{V}$ is a real vector space and subsequently $\mathbf{V}$ has a natural manifold structure \cite{Lee}.
\end{remark}
\begin{remark}[4]
$[\partial_t^k f^n(t)]_{k \in \mathbb{Z}^+, n \in \mathbb{Z}^+, n>1}$ is a basis of a subspace of the Schwartz space $\mathbf{S}(\mathbb{R}^-)$ iff for all $(k,j)$ in $\mathbb{Z}^+ \times \mathbb{Z}^+$ and $k \neq j$, $\partial_t^k f^n \neq \partial_t^j f^n$; and for all $(n_1, n_2)$ in $\mathbb{Z}^+ \times \mathbb{Z}^+$ ($n_1>1$, $n_2>1$),  $\partial_t^k f^{n_1} \neq \partial_t^k f^{n_2}$, $k\in \mathbb{Z}^+$.

\end{remark}
\begin{remark}[5]
We assume  in the follwoing sections that the energy function $\mathcal{E} (F)$ ($F$ in $\mathcal{S}(\mathbb{R})$) is analytic and the convergence of its associated Taylor series development. The convergence property is directly related to the property of $F$ finite energy function which is shown in \textit{Proposition $1$} in \cite{JPMontillet2014}.
\end{remark}
$\mathbf{Definition}$ $3$: \emph{ For $g$ in $\mathcal{C}(\mathbb{R})$, $\exists$ $f^n$ in $\mathcal{S}(\mathbb{R})$, with $f^n(t)$ $\neq$ $0$ and $g(t)$ $\neq$ $0$ for all $t$ in $\mathbb{R}$, we say that the received signal $r(t)$ can be projected onto the vector space defined by the basis $[\partial_t^k f^n(t)]_{k \in \mathbb{Z}^+,n \in \mathbb{Z}^+, n>1}$, if it satisfies the minization condition:}
\begin{eqnarray}\label{receivedsignaldefBISSS}
\min_{k,\beta_{k(l)}}{J(l)}, & & \forall l \in \mathbb{Z}^+ \nonumber \\
\min_{k,\beta_{k(l)}}{ || \sum_{l \in \mathbb{Z}^+} \sum_{k \in \mathbb{Z}^+}\sum_{n \in \mathbb{Z}^+, n>1} \beta_k(l) \partial_t^k f^n(t-\tau_l)-\sum_{l \in \mathbb{Z}^+} a_l(t)  e^{j \phi_l (t)} g(t-\tau_l)|| } & & \nonumber \\
\end{eqnarray}
%
%\emph{if and only if $J(l)$ is convex for 	all $l$}.
%
\begin{discussion}
Definition $3$ can be defined as an optimization problem. As such the optimum solution is found when the cost function $J(l)$ is convex \cite{Wenyu2006}. Depending on the degree of convexity, the optimum solution is either a local  or a global minimizer \cite{Wenyu2006}. The choice of the basis $[\partial_t^k f^n(t)]_{k \in \mathbb{Z}^+,n \in \mathbb{Z}^+, n>1}$ has to be judicious in order to obtain the required convexity property of $J(l)$. Note that $\mathcal{C}(\mathbb{R})$ is the space of continuous functions from $\mathbb{R} \rightarrow \mathbb{R}$ and with their first and second derivative also continuous.
\end{discussion}
%\vspace{0.5em}
%
%
\begin{remark}[6] Following the above discussion, $g$ needs to be in $\mathcal{C}^2(\mathbb{R})$ for an optimization problem, because the convexity property of the cost function $J(l)$ ($l$ in $\mathbb{Z}^+$) is generally established studying its first and second derivative (respectively the Jacobian and Hessian in matrix notation) \cite{Wenyu2006}.
\end{remark}
%Now, comparing the equations \eqref{ImpulseChanneldef} and \eqref{receivedsignaldef}, one can write for $l$ in $\mathbb{Z}^+$.
%%
%\begin{eqnarray}
% \rho_l (t) \partial_t^l f(t-\tau_l)  & \simeq &  a_l(t)  e^{j\phi_l(t)} f(t-\tau_l) \nonumber \\
%\end{eqnarray}
%%
% $f$ follows the conditions exposed in $Property$ $1$. It then exists  $l_0\in \mathbb{Z}^+$ such as for all  $l\in \mathbb{Z}$, $l>l_0$, $\partial_t^l f \sim 0$. Thus,   for all  $l\in \mathbb{Z}$, $l>l_0$, $ a_l(t) \sim 0$. %This remark does not need the assumption of $f$ Wide Sense Stationary, which is used in the appendix (see  Equation \eqref{equationn001b}) on the Taylor Series development.
%
%\vspace{0.5em}
%
$\mathbf{Numerical}$ $\mathbf{example}$ -One can wonder if $\mathcal{S}(\mathbb{R})$ is reduced to $\{\oslash \}$. The trivial answer is to consider the signal $f(t)=0$ (for all $t$ in $\mathbb{Z}^+$). This signal verifies all the properties established in $Property$ $1$, but it is no used in $Definition$ $3$, and not much interest in terms of signal processing applications. However, one can consider an interesting group of signals of  the form $f^n(t) =\exp{(-nt^d)}$ ($d$ in $\mathbb{Z}^+$, $n$ in $\mathbb{Z}^+$, $n>0$). In particular, the family of damped exponentials is defined such as $f^n(t)=  \exp{(\frac{-nt}{\tau})}$ ($n$ in $\mathbb{Z}^+$, $\tau$ in $\mathbb{R}-\{0\}$) \cite{Vanassche}. $f$ is in $\mathcal{S}(\mathbb{R})$ if and only if:
\begin{itemize}
\item[1] $\{ \exists \hspace{0.5em} k_0 \hspace{0.5em} |\hspace{0.5em} \forall \hspace{0.5em} k>k_0, \hspace{0.5em}  \partial_t^k f^n(t) \hspace{0.5em} \sim 0 \hspace{0.5em} \Leftrightarrow  \hspace{0.5em} \tau^k >> n^k \}$ 
\item[2]  $[1] \rightarrow [2]$
\item[3] $\mathcal{E}(f^n)= \int_0^{\infty} |f^n(t)|^2 dt = (\frac{\tau}{2n})$. 
\item[4] Following \cite{Kreizig2003}, the Taylor series development  is $f^n(t) = \sum_{m=0}^\infty \frac{(-nt/\tau)^m}{m!}$ and convergent for $n$ in $\mathbb{Z}^+$ and $\tau$ in $\mathbb{R}-\{0\}$.
\end{itemize}
In addition, one can see that the basis of the vector space describes in $Definition$ $2$ ($[\partial_t^k f^n(t)]_{k \in \mathbb{Z}^+}$) is of dimension $1$, because for $i$ and $j$ in
 $\mathbb{Z}^+$, $\partial_t^i f^n(t)$ and $\partial_t^j f^n(t)$ are collinear. In addition for $n_1$ and $n_2$ in  $\mathbb{Z}^+$ ($n_1>0$, $n_2>0$),$\partial_t^i f^{n_1}(t)$ and $\partial_t^i f^{n_2}(t)$ are also collinear. Using this family, one possible example is to minimize Equation \eqref{receivedsignaldefBISSS} when the channel response is the particular case of the Saleh-Valenzuela indoor propagation channels \cite{Saleh}.
% Now, let us apply the criterion of minimization ...
%
%Another type of decaying exponential signal is $g^n(t)=  \exp{(-nt^2)}$ ($n$ in $\mathbb{Z}^+$). However, it is restricted for the case $t >>1$. Thus, $g$ is in $\mathcal{S}(\mathbb{R})$ if and only if:
%%
%\begin{itemize}
%\item[1] For $t>>1$, $\partial_t^k g^n(t) \sim (-2nt)^k \exp{(-n t^2)} $.  For $k$ in $\mathbb{Z}^+$, $n>0$, $lim_{t \rightarrow \infty}\partial_t^{k} g^n(t) = 0$
%\item[2]  $[1] \rightarrow [2]$
%\item[3] $\mathcal{E}(g^n) = \int_0^{\infty} |g^n(t)|^2 dt \simeq \frac{1}{2} $ for $n>>1$.
%\end{itemize}
%%
%Note that $\mathcal{E}(g^n)$ is always bounded for $n$ in $\mathbb{Z}^+$ \cite{Kreizig2003}.
%%
%Finally,  these signals should follow the rules of the special channel impulse response if $Property$ $1$ can be applied. 
%
\subsection{Multipath fading and energy operators}
\begin{figure}[htb!]
\centering
 \includegraphics[width=0.8\textwidth]{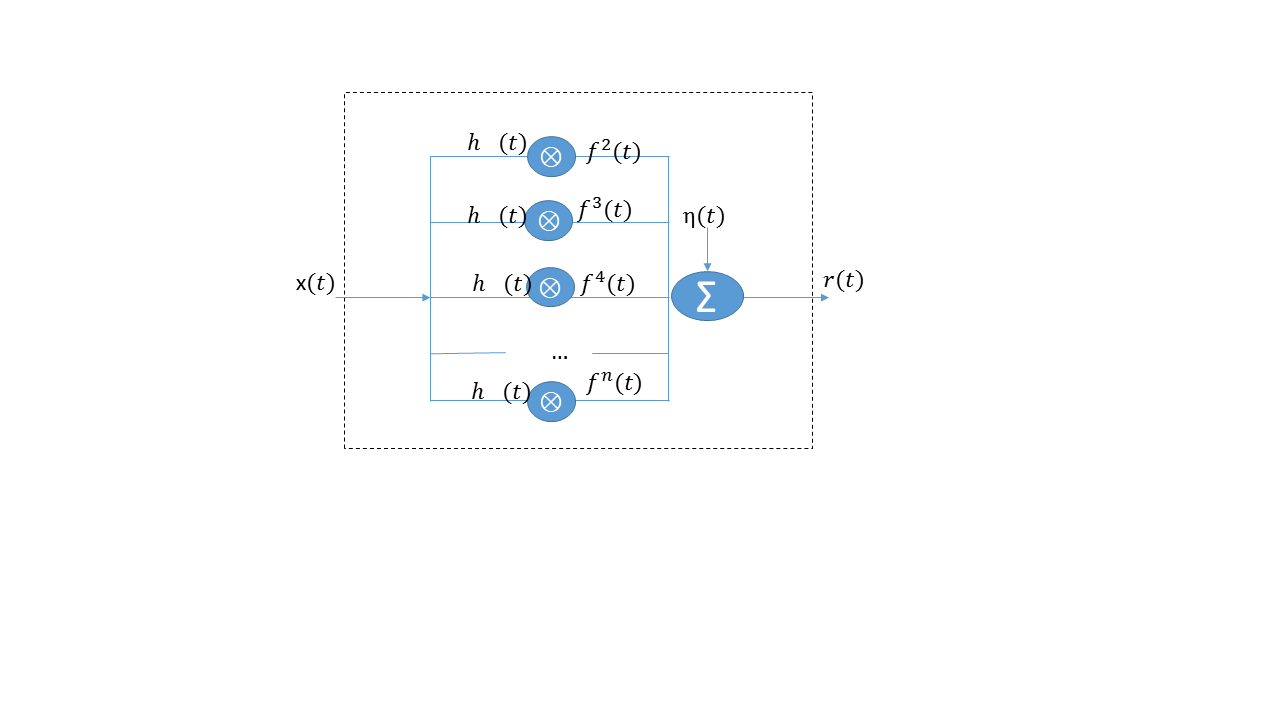}
 \caption{Proposed Channel model as a decomposition in multiple subchannels.}
 \label{Figure10}
\end{figure}

Now, let us consider $f^n$ in $\mathcal{S}(\mathbb{R})$ ($\mathcal{S}(\mathbb{R})\subsetneq \mathbf{S}^-(\mathbb{R})$), $n$ in $\mathbb{Z}^+$ and $n >1$. One can use $Lemma$ $0$  in order to decompose $\partial_t^l f^n$. It then exists $\alpha_n \in \mathbb{R}$ and $\alpha_n \neq 0$ such as  $\partial_t^l f^n = \alpha_n \partial_t^{l-1}(f^{n-2} \Psi_1^+(f))$. Now, if $f^n(t)$ (for all $t$ $\in$ $\mathbb{Z}^+$) is a signal transmitted through $h(t)$, and $r(t)$ the associated received signal. Equation \eqref{receivedsignaldef} can be written such as:
\begin{eqnarray}\label{receivedsignaldefBIS}
r(t) &=& \sum_{n \in \mathbb{Z}^+, n>1} \sum_{l \in \mathbb{Z}^+} \rho_l(t) \partial_t^l f^n(t-\tau_l) + \eta(t) \nonumber \\
&=& \sum_{n \in \mathbb{Z}^+, n>1} \sum_{l \in \mathbb{Z}^+} \rho_l(t) \alpha_n \partial_t^{l-1}(f^{n-2}(t-\tau_l)  \Psi_1^+(f(t-\tau_l) ))+ \eta(t) \nonumber \\
&=&  \sum_{n \in \mathbb{Z}^+, n>1} h(t) \otimes f^n(t) + \eta(t)
\end{eqnarray}
$h$ can be described as a "subchannel" transfer function such as $h(t) = \sum_{l \in \mathbb{Z}^+} \sum_{k \in \mathbb{Z}^+} \beta_k(l) \partial_t^k \delta (t-\tau_l) $
%We did not choose $f^n$ in the subset  $\mathbf{s}^{-}(\mathbb{R})$ $\subsetneq \mathbf{S}^{-}(\mathbb{R})$ in order to apply $\bold{Theorem}$ $0$ \cite{JPMontillet2013}. This restriction plays a certain role with energy spaces (e.g. \cite{JPMontillet2014}), but  it is not justified in this signal processing application.
%
Note that Equation \eqref{receivedsignaldefBIS} emphasizes the key role of the energy operator family $(\Psi_k^+(f))_{k\in\mathbb{Z}}$ in order to determine $l$. In other words, one can estimate $\partial_t^{l-1}\Psi_1^+(f(t-\tau_l) ))$ in order to find $l_0$ such as for all $l> l_0$ $\partial_t^{l}\Psi_1^+(f(t-\tau_l) )) \sim 0$. Looking at Figure \ref{Figure10}, the proposed channel model can be divided in $n$ subchannels with $h$ the "subchannel" transfer function.
Knowing $l_0$, it then limits the computation time restricting $k$ in Equation \eqref{receivedsignaldefBIS}. $n$ can also be restricted with the approximation in Equation \eqref{receivedsignaldefBISSS}, or by using a family of functions (i.e. damped exponentials).
\section{Energy Operators and Matched Filters}\label{section4}
We are now showing the dual relationship between the energy operator family $\big ( \Psi_k^+ \big)_{k\in\mathbb{Z}^+}$ and the SNR with a signal $f^n$ in $\mathcal{S}(\mathbb{R})$. Let us consider the transmitted signal $x(t)+ \nu(t)$. $\nu$ is considered AWGN. From Equation \eqref{receivedsignaldefBIS}, we can write : 
\begin{eqnarray}\label{receivedsignaldefBIS2}
r(t) &=& H(t) \otimes (x(t)+ \nu(t) )\nonumber \\
%r(t) &=& \sum_{n \in \mathbb{Z}^+, n>1} \sum_{l \in \mathbb{Z}^+} \rho_l(t) \partial_t^l f^n(t-\tau_l) + \eta(t) \nonumber \\
%&=& \sum_{n \in \mathbb{Z}^+, n>1} \sum_{l \in \mathbb{Z}^+} \rho_l(t) \alpha_n \partial_t^{l-1}(f^{n-2}(t-\tau_l)  \Psi_1^+(f(t-\tau_l) ))+ \eta(t) \nonumber \\
%
&=&  \sum_{n \in \mathbb{Z}^+, n>1} h(t) \otimes f^n(t) + \sum_{n \in \mathbb{Z}^+, n>1} h(t) \otimes \nu(t) \nonumber \\
r(t)&=& y_s(t) +y_n(t) \nonumber \\
\end{eqnarray}
in matrix notation, 
\begin{eqnarray}\label{equationSNR2}
%
%r(t)&=& f^2(t) \otimes h(t) + \eta(t) \otimes h(t) \nonumber \\
%r(t)&=& y_s(t) +y_n(t) \nonumber \\
%\mathbf{r} &=& \mathbf{h}^H \mathbf{f} \odot \mathbf{f} +\mathbf{h}^H \mathbf{\eta} \nonumber \\
\mathbf{r} &=&  \sum_{n \in \mathbb{Z}^+, n>1} \mathbf{h}^H (\mathbf{f}(n) + \nu)\nonumber \\
\mathbf{r} &=& \mathbf{h}^H \mathbf{F}+\mathbf{h}^H \mathbf{\nu}
\end{eqnarray}
%
%%%%
%\clearpage
%\underline{Case $n =2$}: Following the previous section, let us take the particular case of $f^2$ in  $\mathcal{S}(\mathbb{R})$. The received signal can be written:
%%
%\begin{eqnarray}
%r(t)&=& f^2(t) \otimes h(t) + \eta(t) \otimes h(t) \nonumber \\
%r(t)&=& y_s(t) +y_n(t) \nonumber \\
%\mathbf{r} &=& \mathbf{h}^H \mathbf{f} \odot \mathbf{f} +\mathbf{h}^H \mathbf{\eta} \nonumber \\
%\mathbf{r} &=& \mathbf{h}^H \mathbf{F}+\mathbf{h}^H \mathbf{\eta}
%\end{eqnarray}
%%
%The last line is the vector notation. $\otimes$ is the correlation operator.
 %$\odot$  is a scalar product.
$\mathbf{f}(n)$ is defined as $n$ time a scalar matrix product of the vector $\mathbf{f}$ (i.e. $\mathbf{f}(2) = \mathbf{f}\odot \mathbf{f} $ ). $H$ is the Hermitian transpose operator. From \cite{Proakis}, a matched filter maximized the signal-to-noise ratio:
\begin{eqnarray}
SNR &=& \frac{|y_s|^2}{E\{|y_n|^2\}} \nonumber \\
SNR &=& \frac{|\mathbf{h}^H \mathbf{F} |^2}{E\{|\mathbf{h}^H \mathbf{\nu}|^2\}} \nonumber \\
SNR &=& \frac{|\mathbf{h}^H \mathbf{F} |^2}{\mathbf{h}^H \mathbf{R}_{\nu}\mathbf{h}}
\end{eqnarray}
$\mathbf{R}_{\nu}$ is the correlation matrix of the noise vector $\nu$ defined as $\mathbf{R}_{\nu} = E\{ \mathbf{\nu} \mathbf{\nu}^H\}$. $E\{\}$ is the expectation operator. Following \cite{Proakis}, one can reduce this equation further to :
\begin{eqnarray}
SNR &=& \mathbf{F}^H  \mathbf{R}^{-1}_{\nu} \mathbf{F}
%SNR &=& \mathbf{F}^H  \mathbf{R}^{-1}_{\eta} \mathbf{F}
\end{eqnarray}
But it is also possible to define the SNR for each $\mathbf{f}(n)$ with Equation \eqref{equationSNR2} such as
\begin{eqnarray}
%SNR &=& \frac{|y_s|^2}{E\{|y_n|^2\}} \nonumber \\
SNR_n &=& \frac{|\mathbf{h}^H \mathbf{f}(n) |^2}{E\{|\mathbf{h}^H \mathbf{\nu}|^2\}} \nonumber \\
SNR_n &=& \mathbf{f}(n)^H  \mathbf{R}^{-1}_{\nu} \mathbf{f}(n)
\end{eqnarray}
\section{Discussions}
This work shows the relationship between the energy operators and matched filter when the received signal can be approximated with functions in $\mathcal{S}(\mathbb{R})$ as defined in Equation \eqref{receivedsignaldefBISSS}. The properties of these functions are summed up in $Property$ $1$. This approximation allows to study the received signal as a sum of subchannels with a SNR defined for each of them. The energy operator family $(\Psi_k^+)_{k\in \mathbb{Z}^+}$ can help to reduce the computation time when performing the approximation of the received signal. One possible application is to improve the decorrelation between received signal and multipath fading channel such as in a Rake receiver.
\end{document}